\documentclass[aps,prl,showpacs,twocolumn,groupedaddress]{revtex4}

\usepackage{amsmath,amssymb}
\usepackage{graphicx}
\usepackage{psfrag,color}

\newcommand{\mbf}{\mathbf}

\renewcommand\mark[1]{\bgroup\color{red}\bfseries{[#1]}\egroup}

\begin{document}

\title{{The jamming transition as probed by quasistatic shear flow}}

\author{Claus Heussinger}\author{Jean-Louis
  Barrat}
\affiliation{
 Universit\'e de Lyon; Univ. Lyon I, Laboratoire de
Physique de la Mati\`ere Condens\'ee et Nanostructures; CNRS, UMR
5586, 69622 Villeurbanne, France}
\begin{abstract}
  We study the rheology of amorphous packings of soft, frictionless particles
  close to jamming. Implementing a quasistatic simulation method we generate a
  well defined ensemble of states that directly samples the system at its
  yield-stress. A continuous jamming transition from a freely-flowing state to a
  yield stress situation takes place at a well defined packing fraction, where
  the scaling laws characteristic of isostatic solids are observed. We propose
  that long-range correlations observed below the transition are dominated by
  this isostatic point, while those that are observed above the transition are
  characteristic of dense, disordered elastic media.
\end{abstract}

 \date{\today}

\maketitle

%%--------------------------------------------------------------------------------

% jamminng

A collection of spherical particles, interacting via a finite-range repulsive
(contact) potential, unjams from a solid to a non-rigid state when being
decompressed below a critical volume-fraction,
$\phi_c$~\cite{ohern03,maksePRL1999}. This transition, which has been given the
name ``point J'', is accompanied by several interesting and nontrivial scaling
relations in the solid phase~\cite{ohern03,majmudarPRL2007}. There, pressure and
linear elastic shear modulus vanish as does the ratio of shear to bulk modulus.
The average number of particle contacts jumps from a finite value $z_0$ at point
J to zero just below the transition. The value of $z_0$ is compatible with
Maxwell's estimate for the rigidity transition and signals the fact that at
point J each particle has just enough contacts for a rigid/solid state to exist
({which is called ``isostatic'' state})~\cite{mou98,tkaPRE1999}.  Above the
transition, additional contacts are generated according to the surprisingly
universal law, $\delta z\sim \delta\phi^{1/2}$. While the system moves away from
its isostatic state the effective size of the remaining isostatic regions,
$l^\star\sim \delta z^{-1}$ \cite{wyart05b,wyart05a} has been argued to provide
the diverging length-scale that endows point J with a certain ``criticality''.
Based on these findings, jamming has been regarded as a ``mixed'' transition
that shares properties of both, discontinuous (jump in contact number) and
continuous phase- transitions (diverging
length-scale)~\cite{henkesPRL2005,schwarzEPL2006}.

A different route to approach point J from the fluid phase has been used in the
flow simulations of Refs.~\cite{olssonPRL2007,hatanoJPSJ2008,peyneauPRE2008}.
Several scaling relations have been reported, some of which depend on model
details.  Some others seem to share the universality encountered in the solid
phase.  In contrast to the linear elastic properties in the solid phase, only
little understanding about the flow properties and their relation to nearby
point J has been achieved up to now.

In this Letter, we employ a quasistatic simulation technique that studies the
borderline between fluid and solid state. As we will see, it combines both
aspects, the elasticity of a solid and the flow of a fluid, in one simulation.
This fact allows us to get insight into how point J, and its isostatic state,
affects the flow properties close by. In particular we show that jamming (i.e.
the development of a finite shear stress) as probed by shear flow should be
viewed as a continuous transition, with isostatic effects showing up in the flow
properties primarily below the transition ($\phi<\phi_c$). In contrast, at
volume-fractions above $\phi_c$ the flow is characteristic for amorphous, but
well connected, materials.

\begin{figure}[h]
 \begin{center}
   \includegraphics[width=0.7\columnwidth,]{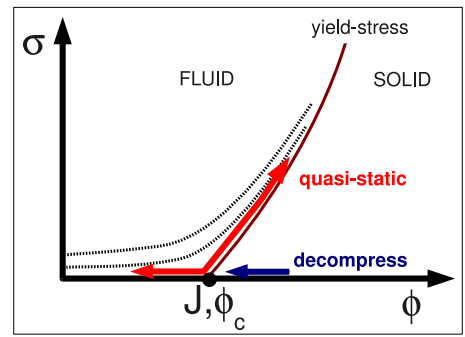}
\end{center}
\caption{Trajectory taken by quasistatic simulations in state-space of shear
  stress $\sigma$ and volume-fraction $\phi$. (Dotted lines) contours of
  constant strain rate $\dot\gamma$. The simulation corresponds to
  $\dot\gamma\to 0$ and thus follows the yield-stress line, $\sigma_y(\phi)$.
  Previous approaches either probe linear elasticity of the solid (decompression
  at $\sigma=0$) or the steady-state flow of the fluid.}
  \label{fig:jamming_diagram}
\end{figure}

The quasistatic simulation probes the flow of the system in the limit of
vanishing shear rate, $\dot\gamma\to 0$ (see Fig.~\ref{fig:jamming_diagram}). On
lowering the volume-fractions from values above $\phi_c$, point J is therefore
approached along the yield-stress line. The resulting ``yield-stress-flow'' is
generated by a succession of equilibrated solid states that, as we will see,
carry the signature of the nearby isostatic state. At point J the yield stress
vanishes such that at lower volume-fractions, $\phi<\phi_c$, the simulation
follows the $\phi$-axis and the system flows at zero stress. This corresponds to
the limiting case, $\dot\gamma\to 0$, of normal fluid flow in the Newtonian
regime~\cite{olssonPRL2007}.

%\section{stress-strain curve}

\begin{figure}[h]
 \begin{center}
   \includegraphics[width=0.9\columnwidth,angle=0]{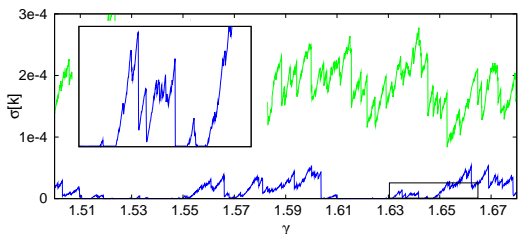}
\end{center}
\caption{Stress-strain relation for a sample of $2500$ particles at
  two different volume fractions, $\phi=0.8470,0.8433$. }
  \label{fig:stress_strain}
\end{figure}

Our system consists of $N$ soft spherical particles that interact with an
harmonic contact interaction with spring constant $k$. As in
Refs.~\cite{ohern03,olssonPRL2007,haxtonPRL2007} we use a $50:50$ mixture of
particles with radius $a$ and $1.4a$ in two dimensions.
%Our system consists of $N$ soft spherical particles with harmonic contact
%interactions as, for example, studied
%in~\cite{ohern03,olssonPRL2007,haxtonPRL2007}.  The mixture consists of two
%types of particles ($50:50$) with size-ratio $1.4$ in two-dimensions. The unit
%of length is the diameter, $a$, of the smaller particle, the unit of energy is
%$ka^2$, where $k$ is the spring constant of the interaction potential. 
In the
simulations the volume fraction $\phi$ is controlled, while stresses are allowed
to fluctuate. Primary output are the pressure, $p$, and (shear-)stress $\sigma$
as a function of imposed strain, $\gamma$ (see Fig.~\ref{fig:stress_strain}). To
implement the shear, variable Lee-Edwards boundary conditions are used.  The
simulation proceeds by minimizing the total potential energy (using conjugate
gradient techniques~\cite{lammps}) after each affine change in the boundary
conditions and particle coordinates, ($\Delta\gamma=5\cdot10^{-5}$). Thus, as
the energy landscape evolves under shear the system always remains at a local
energy minimum, with all forces fully equilibrated.

As can be seen from Fig.~\ref{fig:stress_strain} a typical feature of
quasistatic stress-strain relations is the presence of elastic branches, where
the stress grows linearly with strain. This reversible elastic loading is
terminated by irreversible plastic events during which the stress drops rapidly
and energy is dissipated. The succession of elastic and plastic events defines
the flow of the material just above its yield-stress $\sigma_y(\phi)$. On
lowering the volume-fraction the average stress-level decreases until, below
$\phi_c$, it vanishes and the system flows at zero stress. For intermediate
volume-fractions, close to $\phi_c$, one infers from
Fig.~\ref{fig:stress_strain} that the stress-signal is highly intermittent
showing a coexistence between { the two states of yield-stress and zero-stress
  Newtonian flow.} Note, the striking similarity with the stress-strain curves
of the experiments of Behringer {\it et al.} (see e.g. \cite{behringerPRL2008}).

%\section{coexistence of jammed and flowing phase}

\begin{figure}[h]
 \begin{center}
   \psfrag{1/L}{\tiny $1/\sqrt{N}$}
    \includegraphics[width=0.9\columnwidth,angle=0]{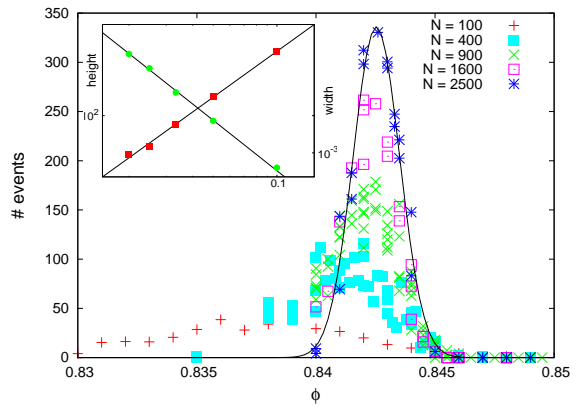}
    \end{center}
    \caption{Number of jamming events (defined as states where stress first
      exceeds a threshold value, $\sigma_{\rm thresh}=2e-07$) %leading from a
                                                              %flowing to a
                                                              %jammed state
      as function of volume fraction $\phi$ and system size $N$.  Inset:
      Finite-size scaling indicates a vanishing width and a diverging height as
      $N\to\infty$.}
  \label{fig:nrJFEvents}
\end{figure}

We have counted the number of jamming-events that lead from the zero-stress flow
to the yield-stress flow at finite stress, and back (see
Fig.~\ref{fig:nrJFEvents}). At high volume fractions the system always flows at
finite stress so no event occurs. The same is true at low volume fractions where
the system always flows at zero stress. Thus, we expect a maximum at an
intermediate volume fraction $\phi_c$ where the system is highly unstable and
makes frequent transitions between the two types of flows. For increasing system
size we find the width of the coexistence region to decrease, presumably
vanishing in the thermodynamic limit. The extrapolated peak position,
$\phi_c(\infty)=0.8433\pm0.0003$, can thus be used to define a critical density.
At this volume-fraction the system develops a yield-stress and the flow changes
from being ``liquid-like'' to ``solid-like''.

The results obtained here are similar to the finite-size scaling in
Ref.~\cite{ohern03}. There is an important difference, however. In that study
statistical information is gathered by generating initial states that are as
random as possible. The observable is the probability that such a randomly
chosen state is jammed.  In contrast, here we are concerned with states that are
connected by the trajectory of the system itself. We are quantifying a dynamical
transition rate at which the system, in the course of shear, undergoes a jamming
transition.

%\section{elastic properties in jammed phase}

One may view the equilibrated finite-stress states visited during the flow
($\phi>\phi_c$) as an ensemble of jammed states that is generated, not by an
external protocol, but by the energy landscape itself. In this spirit we have
analyzed the elastic branches and found good agreement with previously
identified anomalous scaling properties characteristic of marginally rigid
solids. In particular, we found the shear modulus $g$ to scale with the square
root of pressure, $g\sim p^{1/2}$ (not shown). Note, however, that this shear
modulus is \emph{not} the linear elastic modulus of the system
$g(\sigma=0)\equiv g_{\rm lin}$, as discussed in Ref.~\cite{ohern03}. Rather, it
is defined at (or just below) the yield-stress, $g=g(\sigma=\sigma_y)$. As the
yield-stress itself is pressure dependent (we find $\sigma_y\sim p$) the modulus
may in fact show a complicated dependence on pressure, $p$. This is not the
case, however, if we assume $g$ to take the scaling form,
$g(\sigma,p)=p^{1/2}F(\sigma/\sigma_y(p))$. At the yield stress, the argument of
the scaling function is a constant and we recover $g\sim p^{1/2}$ just as at
zero stress. Note, that the presence of a scaling function $F$ is to be expected
close to point J. It has furthermore explicitly been evidenced in the elastic
network model of Ref.~\cite{wyartPRL2008}.

\begin{figure}[ht]
 \begin{center}
   \psfrag{1/L}[l]{\tiny $1/\sqrt{N}$}
    \includegraphics[width=0.9\columnwidth,angle=0]{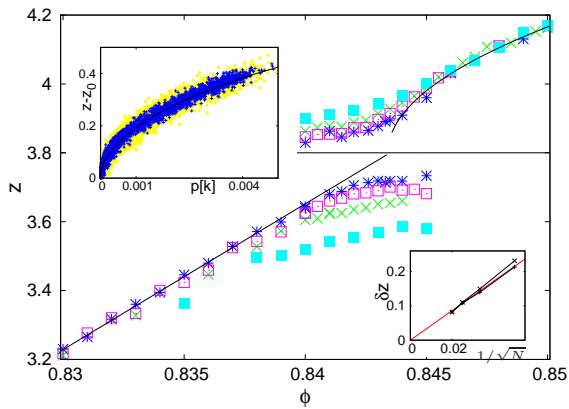}
    \end{center}
    \caption{Contact number $z$ as function of pressure (left inset) and
      volume-fraction $\phi$ (main panel). Symbols are as in
        Fig.~\ref{fig:nrJFEvents}. Right inset: scaling of gap width,
      $\delta z_{gap}\equiv 3.8-z$, with system-size for
        $\phi=0.844,0.843$.
      %two different volume-fractions.
      }
  \label{fig:nr_contacts}
\end{figure}

% inequality vs. equality
Another hallmark of marginally rigid solids is the scaling of the contact
number, $z$. In agreement with the scenario under decompression we find, $\delta
z \equiv z-z_0 \sim p^{1/2}$ (see Fig.~\ref{fig:nr_contacts}). The number of
contacts at zero pressure, $z_0\approx 3.8$, is slightly smaller than the
isostatic value, as rattlers have not been accounted for~\footnote{We have also
  performed simulations with rattlers defined as particles with zero contacts.
  The resulting estimate for $z_0$ is much closer to the expected value of 4.}.
The scaling has been shown to result from a competition between two terms in an
expansion of the elastic energy, $W \sim k u_\parallel^2 -
pu_\perp^2$~\cite{EllenbroekPRL2006}.  In this expression, $u_\parallel$ denotes
the change in compression of a particle contact, while $u_\perp$ relates to the
rotation of the two particles around each other. It has been shown
(\cite{wyart05a,wyart05b}) that both displacement components are related by
$u_\parallel \sim u_\perp\delta z$, such that the energy can be written as,
$W/u_\perp^2\sim k\delta z^2-p$. Thus, both terms are of the same size, when
$\delta z^2\sim p$, which is the scaling relation observed numerically.

In decompression simulations, the number of contacts displays a discontinuity at
$\phi_c$ and jumps from the isostatic value, $z_0$, to zero. This reflects the
absence of any dynamics that would rearrange the particles once they have lost
contact. In contrast, steady shear as studied with the quasistatic simulation
leads to structural rearrangements and thus to particle contacts even at low
volume fractions below $\phi_c$. As can be seen in Fig.~\ref{fig:nr_contacts},
there is still a gap in the number of contacts between jammed finite-stress
(upper branch) and zero-stress (lower branch) configurations.  However, this is
just a sampling artifact and reflects the tendency for jammed configurations to
have $z>z_0$, while zero-stress states generally have $z<z_0$.  As can be
verified in the inset, the width of the gap is reduced with increasing the
system size and seems to vanish in the thermodynamic limit, where only states at
$z_0$ are sampled.
%sample width $1/L$ which is the scaling in the inset.

Furthermore, we find (see Fig.~\ref{fig:nr_contacts}) that below $\phi_c$ the
number of contacts simply increases \emph{linearly} with volume-fraction,
$\delta z\sim\delta\phi$, in striking contrast with the square-root behavior
found above $\phi_c$. The resulting cusp at $\phi_c$ is illustrated in the
figure. The different scaling above and below suggest that different mechanisms
are responsible for the formation of contacts. Indeed, the energetic competition
that controls the number of contacts in the jammed system is absent below
$\phi_c$, where contact formation is of purely geometric origin (the excluded
volume of the particles). Interestingly, this also implies that the isostatic
length-scale $l^\star\sim |\delta z|^{-1}$ scales differently with
volume-fraction above and below jamming. To define $l^\star$ below $\phi_c$,
note, that an isostatic cluster can be equilibrated by blocking its "surface"
degrees of freedom.  The scale $l^\star$ is then defined (similarly as above
$\phi_c$ \cite{wyart05a,wyart05b}) as the size of clusters, in which the number
of extra degrees of freedom in the bulk (due to missing contacts, $\delta z
{l^\star}^d$) is comparable to the number of surface degrees of freedom,
${l^\star}^{d-1}$.

%msq-disp

To summarize this part, we find the ensemble of solid states at the
yield-stress, $\sigma_y(\phi)$, to display the scaling laws characteristic of
marginally rigid solids. As the contact number does not show a discontinuity at
$\phi_c$ we should view jamming under shear as a conventional continuous
transition, in contrast with the mixed character ascribed to the transition seen
under decompression.

\begin{figure}[h]
 \begin{center}
    \includegraphics[width=0.5\columnwidth,angle=-90]{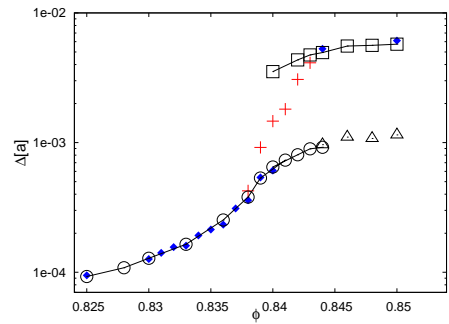}
    \end{center}
    \caption{Non-affine displacements $\Delta$ ($N=900$) taken from: average
      over all configurations (red plus), restricted average over flowing states
      (circles), jammed states (square) and jammed states without plastic events
      (triangles). (Small closed symbols) flowing/jammed states at $N=1600$.}
  \label{fig:msqdisp}
\end{figure}

This continuity is also present on the single-particle level as we will discuss now.
%furthermore evidenced in the single particle displacements
%which we will discuss now. 
For non-interacting particles, the particle displacements are affine,
$\mbf{u}=\gamma x\mbf{\hat e_y}$ (for shear in y-direction).  Interactions lead
to additional non-affine motion, in particular to a non-zero $x$-component,
$u_{\rm na}\equiv \mbf{u}\cdot\mbf{\hat e_x}$.
%which we analyze in the following. 
In Fig.~\ref{fig:msqdisp} we display $\Delta\equiv\langle u_{\rm
  na}^2\rangle^{1/2}$ as a function of volume fraction, $\phi$.

We find perfect continuity of the particle displacements across $\phi_c$
\emph{only} if we exclude the plastic events from the average and consider the
displacements either in the reversible elastic states (triangles) or the states
at zero-stress (circles). Plastic events are, in general, violent rearrangements
that span the whole system. Consequently, these events dominate the amplitude of
the mean-square displacement in the finite-stress states (squares).

{From the continuity of the displacements we conclude that the mechanism that
  leads to nonaffine motion in the ensemble of solid states above $\phi_c$ is
  the same as that in the fluid flow just below $\phi_c$. The difference being
  that above $\phi_c$ particle motion leads to the build-up of stress and its
  subsequent release in the plastic events. Apparently, this does not affect the
  magnitude of the (elastic) non-affine displacements.  However, it \emph{does}
  show up in their correlations.}

%From the continuity of the displacements we conclude that isostatic elasticity
%seen in the elastic states above $\phi_c$ may, in fact, be more relevant for the
%liquid-like flow below $\phi_c$, as there are no plastic events that would
%otherwise dominate the particle dynamics.

%\section{correlation function}

\begin{figure}[h]
 \begin{center}
    \includegraphics[width=0.8\columnwidth,angle=0]{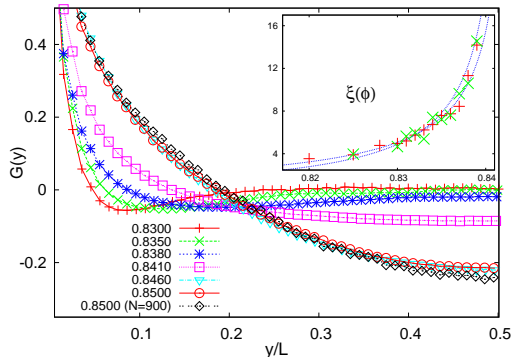}
    \end{center}
    \caption{Correlation function $G(y)$ ($N=2500$) for various volume-fractions
      above and below $\phi_c$. {For $\phi\lessapprox\phi_c$ correlations
        do not decay to zero, probably due to the peridiodicity of the
        simulation box.} Inset: Length-scale $\xi$ as defined by the position of
        the minimum. % increases towards $\phi_c=0.8433$ from below.
        Dashed lines are $\sim\delta\phi^{-0.8}$ and $\sim\delta\phi^{-1}$
        ($\phi_c=0.8433$).}
  \label{fig:correlation}
\end{figure}

%To further test this hypothesis we try to extract a length-scale from our
%simulations. 

Following Ref.~\cite{olssonPRL2007} we study the correlation function
$G(\mbf{r})=\langle u_{\rm na}(\mbf{r}) u_{\rm na}(0) \rangle$ taken at
$\mbf{r}=\mbf{\hat e_y}y$, thus pointing along the shear direction. As in
Ref.~\cite{olssonPRL2007} we find a minimum of the correlation function in the
freely flowing phase ($\phi<\phi_c$). The corresponding length-scale
$\xi(\delta\phi)$ grows with approaching the critical volume fraction from
below. Unfortunately, extracting a reliable value for the correlation length
exponent $\nu$ is hampered by the restricted range of only one order of
magnitude. Values may range from $\nu=0.8\ldots 1.0$ as indicated in the inset
of Fig.~\ref{fig:correlation} (no fit). Note, that a data fit would be highly
sensitive to the value of $\phi_c$ used.  This may also be the reason why
previous studies report smaller values, $0.6\ldots0.7$
~\cite{olssonPRL2007,droccoPRL2005}. With an exponent of $\nu=1$ this may indeed
be the isostatic length-scale, $\xi\equiv l^\star\sim \delta
z^{-1}\sim\delta\phi^{-1}$, thus, indicating the presence of isostatic clusters
that grow on approaching $\phi_c$ from below.

A quite different behavior is found in the regime above $\phi_c$. Here $G(y)$
does not have a minimum and always decays monotonously.  The relevant
length-scale is the system-size, $\xi\sim L$. This is consistent with findings
in Ref.~\cite{maloneyPRL2006} where the lack of observable length-scale is
explained with the dominance of long-range elastic couplings that place
yield-stress flows automatically into a system-size dominated regime.

These results suggest the following picture for the flow at small strain-rates:
the flow properties at small (and zero) stress and volume-fractions below
$\phi_c$ are governed by the vicinity to the isostatic state. This is evidenced
by a correlation length that diverges on approaching the isostatic point J from
below.
%This is evidenced by identifying the correlation length with the isostatic
%length-scale.
As a consequence, one may view the flow as due to the rearrangements of a liquid
of marginally rigid, isostatic clusters. Close to point J a cross-over takes the
system to a different regime ($\phi>\phi_c$), where the flow properties are
dominated by long-range elastic interactions. This regime, thus, reflects the
physics of amorphous, but well connected, materials, where flow is due to the
irreversible rearrangements of liquid-like defects in a solid matrix.

%~\cite{PicardPRE2005}.

%, as for example described in
%the phenomenological model of Ref.

\begin{acknowledgments}
  The authors acknowledge fruitful discussions with Pinaki Chauduri and Ludovic
  Berthier, as well as thank the von-Humboldt Feodor-Lynen, the Marie-Curie
  Eurosim and the ANR Syscom program for financial support.
\end{acknowledgments}

%\bibliography{../jamming}

\end{document}